\documentclass[fleqn,10pt]{wlscirep}
\usepackage[utf8]{inputenc}
\usepackage[T1]{fontenc}
\usepackage{multirow}
\usepackage{xcolor}

\title{SoftDropConnect (SDC) -- Effective and Efficient Quantification of the Network Uncertainty in Deep MR Image Analysis}

\author[1]{Qing Lyu}
\author[3-10,*]{Christopher T. Whitlow}
\author[1-2,*]{Ge Wang}
\affil[1]{Department of Biomedical Engineering, Rensselaer Polytechnic Institute, Troy, NY, USA}
\affil[2]{Biomedical Imaging Center, Center for Biotechnology and Interdisciplinary Studies, Rensselaer Polytechnic Institute, Troy, NY, USA}
\affil[3]{Comprehensive Cancer Center, Wake Forest School of Medicine, Winston-Salem, NC, USA}
\affil[4]{Brain Tumor Center of Excellence, Wake Forest School of Medicine, Winston-Salem, NC, USA}
\affil[5]{Radiology Informatics \& Image Processing Laboratory, Wake Forest School of Medicine, Winston-Salem, NC, USA}
\affil[6]{Department of Radiation Oncology, Wake Forest School of Medicine, Winston-Salem, NC, USA}
\affil[7]{Department of Radiology, Wake Forest School of Medicine, Winston-Salem, NC, USA}
\affil[8]{Department of Neurosurgery, Wake Forest School of Medicine, Winston-Salem, NC, USA}
\affil[9]{Department of Neurology, Wake Forest School of Medicine, Winston-Salem, NC, USA}
\affil[10]{Department of Biomedical Engineering, Wake Forest School of Medicine, Winston-Salem, NC, USA}

\affil[*]{Corresponding author}

\begin{abstract}
Recently, deep learning has achieved remarkable successes in medical image analysis. Although deep neural networks generate clinically important predictions, they have inherent uncertainty. Such uncertainty is a major barrier to report these predictions with confidence. In this paper, we propose a novel yet simple Bayesian inference approach called SoftDropConnect (SDC) to quantify the network uncertainty in medical imaging tasks with gliomas segmentation and metastases classification as initial examples. Our key idea is that during training and testing SDC modulates network parameters continuously so as to allow affected information processing channels still in operation, instead of disabling them as Dropout or DropConnet does. When compared with three popular Bayesian inference methods including Bayes By Backprop, Dropout, and DropConnect, our SDC method (SDC-W after optimization) outperforms the three competing methods with a substantial margin. Quantitatively, our proposed method generates substantial improvements in prediction accuracy (by 3.4\%, 2.5\%, and 6.7\% respectively for whole tumor segmentation in terms of dice score; and by 11.7\%, 3.9\%, and 8.7\% respectively for brain metastases classification) and greatly reduced epistemic and aleatoric uncertainties. Our approach promises to deliver better diagnostic performance and make medical AI imaging more explainable and trustworthy.
\end{abstract}

\begin{document}

\flushbottom
\maketitle
%
%
\thispagestyle{empty}


\section*{Introduction}
Over the past decade, various deep neural networks (DNNs) have been developed and revolutionized our capabilities in different domains. In the field of medical image analysis, DNNs can now generate far better results than conventional feature-based methods in many applications such as nodule detection \cite{gu2018automatic}, tumor segmentation \cite{isensee2020nnu}, and cancer staging \cite{munir2019cancer}. In these cases, neural networks were mostly trained to produce point estimates by optimizing network parameters to minimize objective functions. Normally, after the training stage all network parameters are fixed and kept unchanged in the inference stage. As a result, for the same input the network will produce the same output. For these neural networks, it is inevitable to generate both accurate and inaccurate predictions, but there is no estimation on the reliability of these predictions. This limitation is major for the deployment of deep learning methods in healthcare-critical real-world applications \cite{gawlikowski2021survey}. 

For tumor imaging \cite{leynes2020attenuation, ballestar2020mri, hepp2021uncertainty, tanno2021uncertainty, narnhofer2021bayesian}, estimating the network uncertainty allows clinicians to reject network predictions of high uncertainty and use additional methods such as biopsy and bio-markers to effectively mitigate false predictions. Generally speaking, there are two types of network uncertainties: model or epistemic uncertainty and data or aleatoric uncertainty. Epistemic uncertainty accounts for fuzziness in the model parameters from data insufficiency and model mismatching, while aleatoric uncertainty measures inherently data noise \cite{kendall2017uncertainties}. Properly measuring uncertainty is of critical importance. Researchers tried to use the softmax in the final layer to infer the confidence of classification predictions, and found that networks could generate inaccurate predictions with high softmax-type pseudo possibility. In this case, the softmax probability only reflects the possibility of a given class relative to other classes on a specific input and does not show an overall confidence \cite{mobiny2021dropconnect, gal2016uncertainty}. 

To better estimate the network uncertainty, many methods were proposed in recent years. Based on the number and the nature of the used DNNs, we can classify them into the single deterministic \cite{sensoy2018evidential, malinin2018predictive, raghu2019direct}, Bayesian \cite{gal2016dropout, blundell2015weight, mobiny2021dropconnect}, ensemble \cite{lakshminarayanan2016simple, valdenegro2019deep, wen2020batchensemble}, and test-time augmentation categories \cite{shorten2019survey, wen2020time, wang2018automatic, wang2019aleatoric}. Among these classes, variational Bayesian inference is a commonly used approach. Different from deterministic DNNs, parameters of a Bayesian neural network (BNN) are not fixed but described with a posterior possibility distribution \cite{neal2012bayesian}. Since the estimation of such a posterior possibility distribution is often intractable, researchers proposed variational inference to approximate the posterior possibility distribution as a series of simpler variational distributions \cite{graves2011practical}. To optimize BNNs, Bayes by Backprop (BBB) \cite{blundell2015weight}, local reparameterization \cite{kingma2015variational}, and multiplicative normalising flows \cite{louizos2017multiplicative} are commonly used training methods. Mathematically equivalent approximation to BNNs, Monte Carlo DropOut is another popular method to estimate the network uncertainty, which performs dropout on every network layer in both training and inference stages \cite{gal2016dropout}. As a generalization to Dropout, DropConnect was recently proposed to replace Dropout, which is called Monte Carlo DropConnect \cite{mobiny2021dropconnect}.

In this paper, we modify DropConnet so that weights in drop masks are modulated according to a continuous uniform distribution instead of a discrete Bernoulli distribution, extending the DropConnect method to a generalized version, which is referred to as SoftDropConnect (SDC). Then, we investigate the performance of SDC using Monte Carlo methods for network uncertainty estimation in medical applications by comparing our method with other variational Bayesian inference methods. Our main contributions in this paper are as follows:

\begin{enumerate}
\item SDC is designed as a generalized version of DropConnect by sampling drop weights from a uniform distribution.

\item SDC is applied to estimate the network uncertainty. In comparison with other variational Bayesian inference methods, the utilities and merits of SDC are demonstrated.

\item Two variants of SDC, SDC-Strong and SDC-Weak, are further proposed to estimate the network uncertainty. Our results show that SDC-Weak outperforms SDC and SDC-Strong with higher accuracy and lower uncertainty.

\item In two applications for gliomas segmentation and metastases classification respectively, our approach is effective to improve prediction accuracy and avoid false predictions, showing a great potential of this technique in future clinical applications.
\end{enumerate}

\begin{figure}[ht]
\centering
\includegraphics[width=6in]{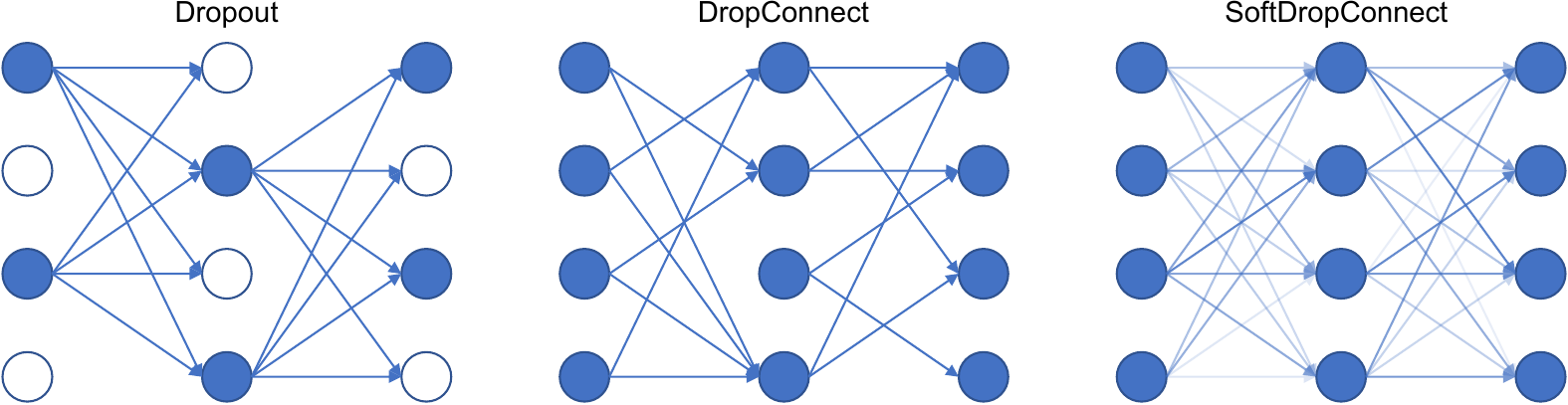}
\caption{Comparison of Dropout (left), DropConnect (middle), and SoftDropConnect (right). For Dropout and DropConnect, connection drop mask weights are either ones (in blue) or zeros (in white). For Soft DropConnect, connection drop mask weights are continuous; for example, between one and zero.}
\label{fig1}
\end{figure}

\section*{Methods}
In this section, we first introduce variational Bayesian inference. Then, we describe neural network structures and experimental designs for all the three tasks performed in this study. Finally, we explain the key metrics used for evaluation of the competing network uncertainty methods.

\subsection*{Variational Bayesian inference}
\subsubsection*{Bayesian neural network}
Importantly, an artificial neural network can be viewed as a probabilistic model $p(y|x,w)$, where $x$, $y$, and $w$ are input, output, and weight parameters respectively. In 2015, Blundell et al. \cite{blundell2015weight} performed an excellent study on weight uncertainty in neural networks. Here we briefly describe their approach using their notations. The parameters of a neural network as a probabilistic model can be optimized via maximum likelihood estimation (MLE) from a training dataset $D=(x_i,y_i)_i$ using a back-propagation algorithm. 
\begin{equation}\label{equ1}
\begin{split}
    w &= \mathop{\arg\max}_{w}p(D|w) \\
    &= \mathop{\arg\max}_{w}\sum_{i}\log p(y_i|x_i,w).
\end{split}
\end{equation}
A more sophisticated way to optimize the probabilistic neural network is called maximum a posterior (MAP), when a parameter prior $p(w)$ is available. In this case, we have
\begin{equation}\label{equ2}
\begin{split}
    w &= \mathop{\arg\max}_{w}p(w|D) \\
    &= \mathop{\arg\max}_{w}\log p(D|w)+\log p(w).
\end{split}
\end{equation}
In the above formulation, the first and second terms reflect the data fidelity and the model complexity respectively.

Based on the parametric prior, a Bayesian neural network (BNN) seeks the posterior distribution of parameters $p(w|D)$, and performs its inference $p(y|x) = \mathbb{E}_{p(w|D)}[p(y|x,w)]$. After being trained, BNN parameters are in terms of possibility distributions, instead of point estimates as in the case of deterministic networks. As a result, network outputs and associated uncertainties can be naturally estimated. The parameter posterior distribution is computed based on
the Bayes theorem:
\begin{equation}\label{equ3}
    p(w|D) = \frac{p(w,D)}{p(D)} = \frac{p(D|w)p(w)}{p(D)}, \\
\end{equation}
where $p(D|w)$ and $p(D)$ are the likelihood function and the probability of the data. As \eqref{equ3} is  intractable, variational inference is introduced to solve the problem. The key idea behind variational inference \cite{graves2011practical, blundell2015weight, gal2015bayesian} is to use a family of variational distributions $q(w|\theta)$ defined by the parameter $\theta$ to approximate the target posterior $p(w|D)$. For example, using a set of Gaussian distributions $\theta\sim N(\mu,\delta^2)$ as a surrogate of $p(w|D)$. The discrepancy should be minimized in terms of the Kullback-Leibler (KL) distance between $p(w|D)$ and $q(w|\theta)$:
\begin{equation}\label{equ4}
\begin{split}
    \theta^{\star} &= \mathop{\arg\min}_{\theta}D_{KL}[q(w|\theta)\|p(w|D)] \\
    &= \mathop{\arg\min}_{\theta}\int q(w|\theta)\log\frac{q(w|\theta)}{p(w)p(D|w)}dw \\
    &= \mathop{\arg\min}_{\theta}D_{KL}[q(w|\theta)\|p(w)] - \mathbb{E}_{q(w|\theta)}[\log p(D|w)].
\end{split}
\end{equation}
Thus, the objective function can be expressed as follows:
\begin{equation}\label{equ5}
    \mathcal{F}(D,\theta) = D_{KL}[q(w|\theta)\|p(w)] - \mathbb{E}_{q(w|\theta)}[\log p(D|w)].
\end{equation}
To solve \eqref{equ5}, Blundell et al. \cite{blundell2015weight} proposed Bayes by Backprop method via Monte Carlo sampling:
\begin{equation}\label{equ6}
    \mathcal{F}(D,\theta) \approx \sum_{i=1}^{n} \log q(w^{(i)}|\theta) - \log p(w^{(i)}) - \log p(D|w^{(i)}),
\end{equation}
where $q(w^{(i)}|\theta)$ denotes the $i^{th}$ Monte Carlo sample drawn from the variational distribution $q(w|\theta)$, and $w^{(i)}$ stands for the $i^{th}$ Monte Carlo sample drawn from the prior distribution $p(w)$. In the training process, the mini-batch gradient descent is computed of
\begin{equation}\label{equ7}
    \mathcal{F}_{i}(D_i,\theta) = \frac{1}{M}D_{KL}[q(w|\theta)\|p(w)]-\mathbb{E}_{q(w|\theta)}[\log p(D_i|w)].
\end{equation}
where $\sum_{i}\mathcal{F}_{i}(D_i,\theta)=\mathcal{F}(D,\theta)$. The dataset is supposed to be uniformly and randomly divided into $M$ minibatches. To solve \eqref{equ7}, Blundell et al. \cite{blundell2015weight} introduced several techniques including unbiased Monte Carlo gradient computation, Gaussian variational posterior, scale mixture prior, KL re-weighting, and Thompson sampling. In this study, we implemented BBB based on \cite{shridhar2019comprehensive}.

\subsubsection*{SoftDropConnect (SDC)}
It is challenging to implement traditional BNNs of complex structures \cite{mobiny2021dropconnect}. Alternatively, Gal et al. showed that Monte Carlo Dropout can be used to perform variational inference \cite{gal2016dropout}. Compared with traditional BNNs, Dropout introduces no additional parameters and can be trained much faster. Additionally, Dropout can also work well with networks of complex structures. Furthermore, Mobiny et al. demonstrated that Monte Carlo DropConnect, replacing Dropout with DropConnect, to approximate Bayesian inference \cite{mobiny2021dropconnect}. It is found that DropConnect extracts uncertainty more robustly \cite{gawlikowski2021survey}. In another study, Dropout and DropConnect was cobmined to produce even higher prediction accuracy with better robustness \cite{mcclure2016robustly}. 

As originally proposed by Wan et al. \cite{wan2013regularization}, DropConnect is a generalization of Dropout \cite{srivastava2014dropout} and is used to regularize neural networks. As shown in Fig. \ref{fig1}, for a fully-connect layers with $M$ input nodes and $N$ output nodes, DropConnect can be formulated as
\begin{equation}\label{equ8}
    v_{out} = \frac{\sigma[z \odot (w \cdot v_{in}]}{1-p},
\end{equation}
where $v_{out}\in\mathbb{N}^{batch \times M}$ and $v_{in}\in\mathbb{N}^{batch \times N}$ are output and input vectors, respectively, $p$ denotes the drop or leave out rate, $w\in\mathbb{N}^{M \times N}$ is the weight matrix, $z\in\mathbb{N}^{M \times N}$ is the drop weight matrix, $\odot$ and $\sigma$ signify the Hadamard product and activation respectively. In this work, we propose a method to change the drop matrix defined by a Bernoulli distribution $z^{ij}\sim Bernoulli(p)$ to the counterpart defined by a continuous uniform distribution $z^{ij}\sim U(0,1)$. Compared with the original DropConnect that connections between nodes in adjacent layers are either kept or totally dropped, our proposed method is gentle, only weakening the connections by multiplying a continuous random weight modulation factor; for example, drawn from a uniform distribution between 0 and 1. To highlight this flexibility, we call our proposed method SoftDropConnect (SDC).

Based on the generic SDC design, we further specialize the SDC into two interesting variants SDC-S and SDC-W
by limiting the random distribution to ranges of interest. Specifically, for SDC-S, $z^{ij}\sim U(0,0.5)$, which weakens connections more strongly than SDC. For SDC-W, $z^{ij}\sim U(0.5,1)$, which modulates network weights more gentle than SDC.

\subsection*{Neural networks}
In each of our selected three tasks, the six variational Bayesian inference methods (BBB, Dropout, DropConnect, SDC, SDC-S, and SDC-W) were used in this study, which have the same network structure design.

\subsubsection*{MNIST digit recognition}

For MNIST digit recognition, we built a simple network with two convolutional blocks followed by two fully-connected layers. Each convolutional block includes two convolutional layers followed by a batch normalization layer and a rectified linear unit (ReLU). Two max-pooling layers (kernel size of 2 and stride of 2) are after each convolutional block. Every convolutional layer has a kernel size of 3, stride of 1, and padding of 1. Totally, 32, and 64 kernels are used for the two convolutional blocks respectively. The first fully-connected layer has 3,136 input nodes and 1,024 output nodes, while the second fully-connected layer has 1,024 input nodes and 20 output nodes.

\subsubsection*{BraTS gliomas segmentation}
We used the nnU-Net \cite{isensee2021nnu} for gliomas segmentation. There are four down-sampling blocks, a bottleneck block, four up-sampling blocks, and a final 3D convolutional layer. In each down-sampling block, there are two 3D convolutional layers. Each convolutional layer is followed by a batch normalization layer and a ReLU layer. 3D max-pooling layers are after each down-sampling block with kernel size of 2 and stride of 2. The number of kernels in down-sampling blocks are 16, 32, 64 and 128 respectively. The bottleneck block has two 3D convolutional layers with 256 kernels followed by batch normalization and ReLU activation. Each up-sampling block has an up-sampling layer with a factor of 2, and three 3D convolutional layers followed by batch normalization and ReLU activation. The number of kernels in the up-sampling blocks are 128, 64, 32 and 16 respectively. Skip connections link the output of the down-sampling blocks and the second convolutional layer in the up-sampling blocks. All the convolutional layers are with a kernel size of 3, stride of 1, and padding of 1.

\subsubsection*{Brain metastases origin site classification}

We modified the tumor classification network used in our previous work \cite{lyu2021transformer} to work on a single GPU. There are two identical feature extraction modules, an attention module, and a fully-connected modules. Compared with the original network, we modified the feature extraction module and the attention module. In the attention module, we changed the number of channel attention convolutional kernels from 512 to 256. In each feature extraction module, there are an initial 3D convolutional layer with 16 kernels and five convolutional blocks with 16, 32, 64, 128 and 256 kernels respectively. Each convolutional block has two identical 3D convolutional layers followed by batch normalization and ReLU activation. There is a 3D maxpooling layer with the kernel size of (1,2,2) and stride of (1,2,2) before the second and fourth convolutional blocks. Before the first, third, and fifth convolutional blocks respectively, there is a 3D maxpooling layer with kernel size of (2,2,2) and stride of (2,2,2). All convolutional layers are with a kernel size of 3, stride of 1, and padding of 1.

\subsubsection*{Implementation details}

For the MNIST digit recognition task, the training process continued for 500 epochs with a learning rate of 0.001 and a batch size of 64. For the BBB method, each image was repeatedly input into the network five times before the gradient back propagation, and other methods only input each image once. For all the methods, each inference was based on repeatedly inputting an image into the network 25 times in each validation step, and 100 times in each testing step. Cross-entropy was used in the objective function. 

When performing the BraTS gliomas segmentation, we set the total number of epochs to 250, a learning rate of 0.001, and a batch size of 2. For the BBB method, different from the MNIST digital classification, the number of input repetitions was 3 instead of 5 in each training gradient back propagation step. At the validation and testing stages, for all the methods each case was repeatedly inferred 10 times and 100 times, respectively. The summation of dice score and cross-entropy served as the objective function.

Similar to the design for gliomas segmentation, the training process for brain metastases tumor origin site classification continued for 250 epochs with a batch size of 2. The learning rate was set to 0.0001. Each case was repeated input to the network 3 times during the BBB training. For all the methods, the protocols for validation and testing were the same as that for gliomas segmentation. For this task, we adopted the weighted cross-entropy as the objective function.

In all the experiment, the Adadelta optimizer \cite{zeiler2012adadelta} was used with a rho of 0.9, eps of 1e-06, and no weight decay. In each epoch, we recorded mean validation accuracy, as shown in Fig. \ref{fig2}a, \ref{fig3}a, and \ref{fig5}a. All the experiments were conducted in PyTorch 1.7.0 on a single NVIDIA Tesla V100 GPU with 32GB memory.

\subsection*{Network uncertainty metrics}
\subsubsection*{Aleatoric uncertainty}
To compute aleatoric uncertainty, we adopted the method proposed by Kendall \cite{kendall2017uncertainties}. Network outputs were split into two halves: the first for prediction and the second for aleatoric uncertainty estimation. At the training stage, the network can be optimized by minimizing the normalized negative log-likelihood function:
\begin{equation}\label{equ9}
    \mathcal{L} = \frac{1}{N}\sum_{i}\frac{1}{2}\exp(-s_i)\|y_i-\hat{y}_i\|^2+\frac{1}{2}s_i,
\end{equation}
where the first term is for data fidelity, the second term for data uncertainty regularization, $\hat{y}_i$ is the first half of the network outputs, and $s_i:=\log\hat{\sigma}_i^2$ is the second half of the outputs.

\subsubsection*{Epistemic uncertainty}
Mutual information (MI) is widely used for measuring epistemic uncertainty for classification tasks\cite{gawlikowski2021survey, smith2018understanding}. MI measures the expected divergence between the stochastic softmax output and the expected softmax output. The expected softmax output can be calculated using as follows:
\begin{equation}\label{equ10}
    \hat{p} = \mathbb{E}_{\theta \sim p(\theta|D)}[p(y|x,\theta)].
\end{equation}
MI uses the entropy to measure the mutual dependence between two variables:
\begin{equation}\label{equ11}
    MI(\theta,y|x,D) = H[\hat{p}]-\mathbb{E}_{\theta \sim p(\theta|D)}H[p(y|x,\theta)],
\end{equation}
where $H$ is the entropy
\begin{equation}\label{equ12}
    H(p) = -\sum_{k=1}^{K}p_{k} \log_{2}(p_{k}).
\end{equation}
In all our three tasks, two of them are for classification, and the other is for segmentation. As segmentation can be seen as pixel-wise classification, we adopted MI to measure epistemic uncertainty in all three tasks. To compute epistemic uncertainty, Monte Carlo method was utilized based on the trained network and MI was calculated.

\section*{Results}
In this study, to estimate the network uncertainty, we propose a SoftDropConnect approach called SDC by replacing drop weights from a Bernoulli distribution into that from a continuous uniform distribution such as $\sim U(0,1)$. As two variants, we limit the drop weight distribution ranges to have SDC-Strong or SDC-S (drop weights $\sim U(0,0.5)$) and SDC-Weak or SDC-W (drop weights $\sim U(0.5,1)$). To show the merits of our approach, we compare the three versions (SDC, SDC-S. SDC-W) of our approach with popular variational Bayesian inference methods: Bayes by Backprop \cite{blundell2015weight} (BBB), Monte Carlo Dropout \cite{gal2016dropout}, and Monte Carlo DropConnect \cite{mobiny2021dropconnect}. For simplicity, we call Monte Carlo Dropout and Monte Carlo DropConnect as Dropout and DropConnect respectively in this paper. Experiments are conducted on three datasets: the MNIST dataset for digit recongnition, the BraTS dataset for gliomas segmentation, and our in-house brain metastases dataset for tumor origin site classification.

\begin{figure}[t]
\centering
\includegraphics[width=\linewidth]{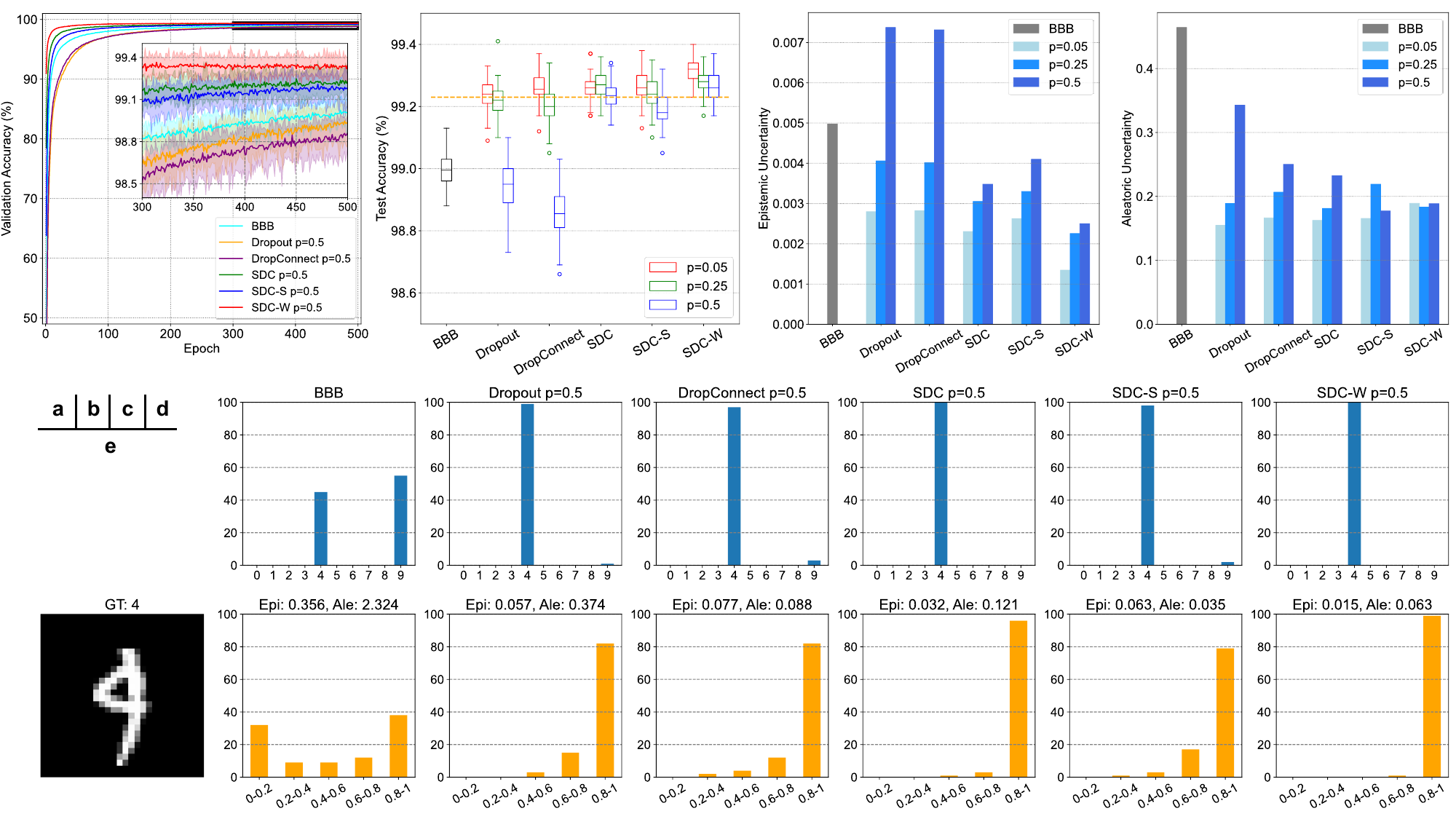}
\caption{Uncertainty estimation results on the MNIST dataset for digit recognition. a) Validation accuracy during the training process. Solid lines show average accuracy and bands reflect the min-max range; b) Boxplot comparison of accuracy results on test data from different methods, where the orange dashed line is the result of the deterministic network with the same structure; c-d) Bar charts of epistemic and aleatoric uncertainty; e) A testing example with a leave out rate $p=0.5$. In e), The top row gives the each class prediction counts, and the subplot title shows the most popular prediction class voting result; The bottom row lists softmax probability counts corresponding to the popular voting class, and the subplot title is the value of epistemic and aleatoric uncertainty.}
\label{fig2}
\end{figure}
\subsection*{Datasets}
\begin{figure}[t]
\centering
\includegraphics[width=\linewidth]{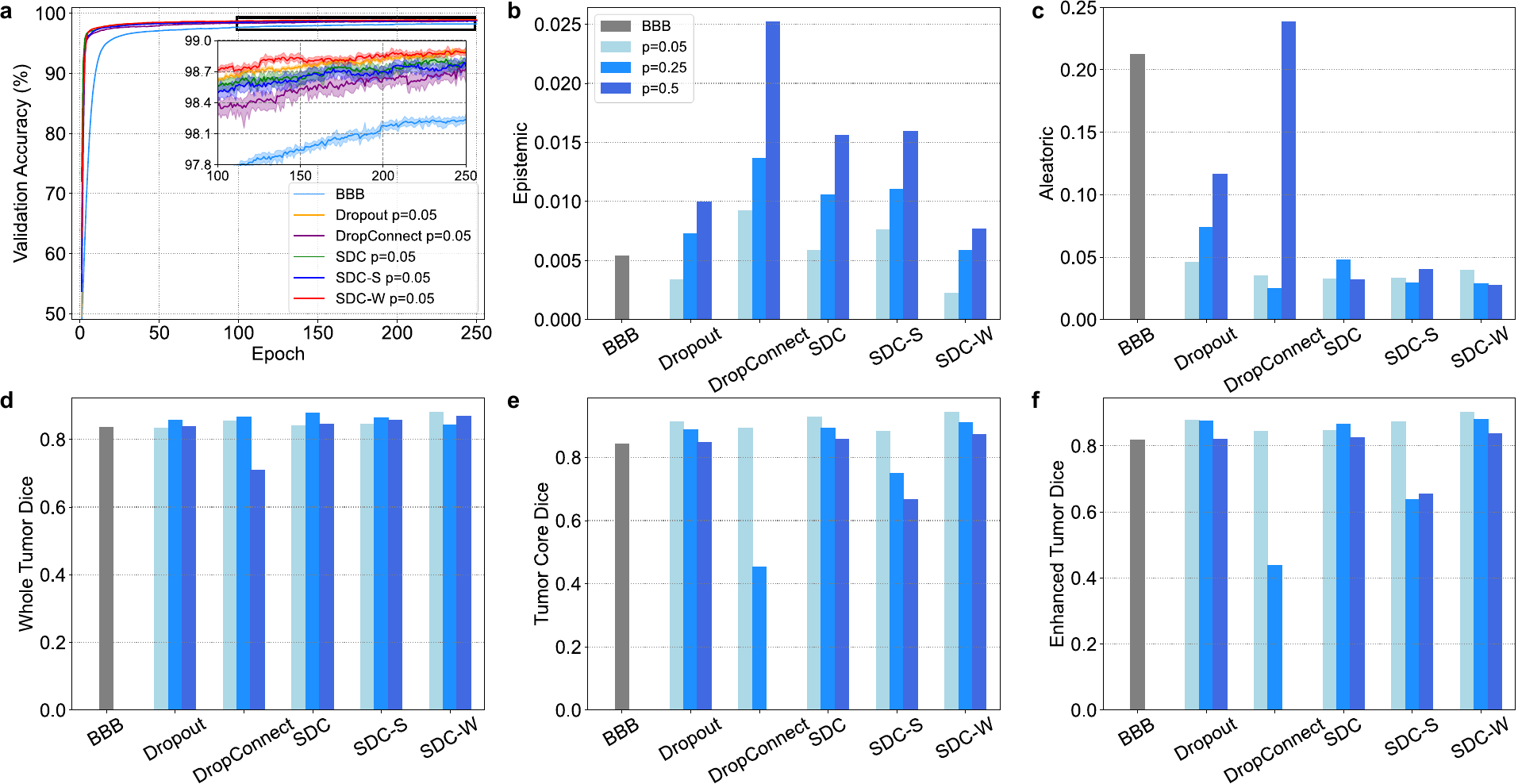}
\caption{Uncertainty estimation results on the BraST dataset for gliomas segmentation. a) Validation accuracy in the training process with solid lines for average accuracy and bands for accuracy min-max range; b-c) Bar charts of epistemic and aleatoric uncertainty; d-f) Bar charts of whole tumor, tumor core, and enhanced tumor dice score from different methods.}
\label{fig3}
\end{figure}

As a starting point, MNIST dataset \cite{lecun1998mnist} was used for digit recognition. In this study, 50,000 images were used for training, and 10,000 images for validation and testing. Then, BraTS 2021 dataset \cite{baid2021rsna} was used for gliomas segmentation. In the provided training dataset, there are 1,251 cases, each of which contains human-labeled segmentation ground truth and MR images in four modalities: T1-weighted, T1 contrast enhancing (T1CE), T2-weighted, and T2 Fluid Attenuated Inversion Recovery (FLAIR). We used the first 400 cases for training, the next 100 cases for validation, and the last 50 cases for testing. Finally, our own brain metastases dataset was used for tumor origin site classification. This dataset was collected at Wake Forest School of Medicine from 2000 to 2021. Detailed information and data pre-processing steps were described in \cite{lyu2021transformer}. As shown in Fig \ref{fig3}a, in this study we used 1,200 cases for training, 150 cases for validation ,and 232 cases for testing.

\subsection*{Uncertain estimation on the MNIST dataset for digit recognition}
The MNIST dataset is widely used as a toy example in machine learning research. In this study, we first used MNIST to demonstrate the effectiveness of our proposed methods for uncertainty estimation. To investigate the network uncertainty, apart from epistemic and aleatoric uncertainties, we also statistically analyzed the prediction results and corresponding softmax possibilities by taking into account the class of predictions and the possibility of the popular voting class.

We compared six variational Bayesian inference methods on the MNIST dataset. As shown in Fig. \ref{fig2}a-d, SDC outperforms BBB, Dropout, and DropConnect with higher accuracy. In terms of epistemic and aleatoric uncertainties, our SDC methods are also advantageous with lower uncertainty levels when comparing with BBB, Dropout, and DropConnect. Among our proposed SDC, SDC-S, and SDC-W methods, SDC-W shows the best performance with the highest accuracy and lowest uncertainty. On average, SDC-W can achieve the highest accuracy, overcoming the deterministic model, BBB, Dropout, DropConnect, SDC, and SDC-S by 0.17\%, 1.24\%, 0.98\%, 0.95\%, 0.20\%, and 0.32\% respectively.

A representative testing case is demonstrated in Fig. \ref{fig2}e. It can be observed that almost every SDC-W prediction is correct, and these predictions are with very high softmax possibilities. Moreover, the SDC-W prediction is with a low level of uncertainty when compared with the other methods. All these results demonstrate that SDC-W is highly confident in its prediction. In contrast, the other results are with more uncertainties. For example, the majority of BBB predictions are counted in the bin for digit 9 instead of 4, showing false predictions. It is seen in the prediction class softmax possibility chart that only about 40\% of all predictions are counted in the last bin with a very high softmax possibility. More examples are presented in Supplementary Fig. 1.

\subsection*{Uncertain estimation on the BraTS dataset for gliomas segmentation}
To demonstrate the relevance of uncertainty estimation to medical applications, we selected the BraTS dataset for gliomas segmentation. We used 1) dice scores of whole tumor, tumor core, and enhanced tumor, 2) absolute error between mean prediction and ground truth, 3) pixel-wise epistemic uncertainty, and 4) pixel-wise aleatoric uncertainty to reflect the segmentation uncertainty.

Similar to our results on the MNIST datasest, it is observed in Fig. \ref{fig3}a that the SDC results are better than the BBB, Dropout, and DropConnect counterparts in terms of validation accuracy. Among SDC, SDC-S, and SDC-W results, the SDC-W results show the highest validation accuracy. In terms of uncertainty, as shown in Fig. \ref{fig3}b-c, both epistemic and aleatoric uncertainty values are the lowest for SDC-W. As far as the effect of leave out rate $p$ on the network performance is concerned, it is shown in Fig. \ref{fig3}d-f that tumor segmentation dice scores of SDC and SDC-W are decreased relatively less than the other methods when the leave out rate $p$ was increased from 0.05 to 0.5. DropConnect, on the other hand, is heavily affected by the change of the leave out rate as the segmentation results of tumor core and enhanced tumor degrade quickly when the leave out rate is increased. When $p$ reaches 0.5, there is hardly no tumor core and enhanced tumor segmented. Quantitatively, compared with BBB, Dropout, DropConnect, SDC, and SDC-S, SDC-W significantly lower epistemic uncertainty by 2\%, 23\%, 67\%, 50\%, and 54\% respectively, and reduce aleatoric uncertainty by 85\%, 60\%, 68\%, 15\%, and 7\% respectively. For the whole tumor dice score, SDC-W presented results indicate the improvements by 3.4\%, 2.5\%, 6.7\%, 1.1\%, and 0.9\%, respectively.

\begin{figure}[t]
\centering
\includegraphics[width=\linewidth]{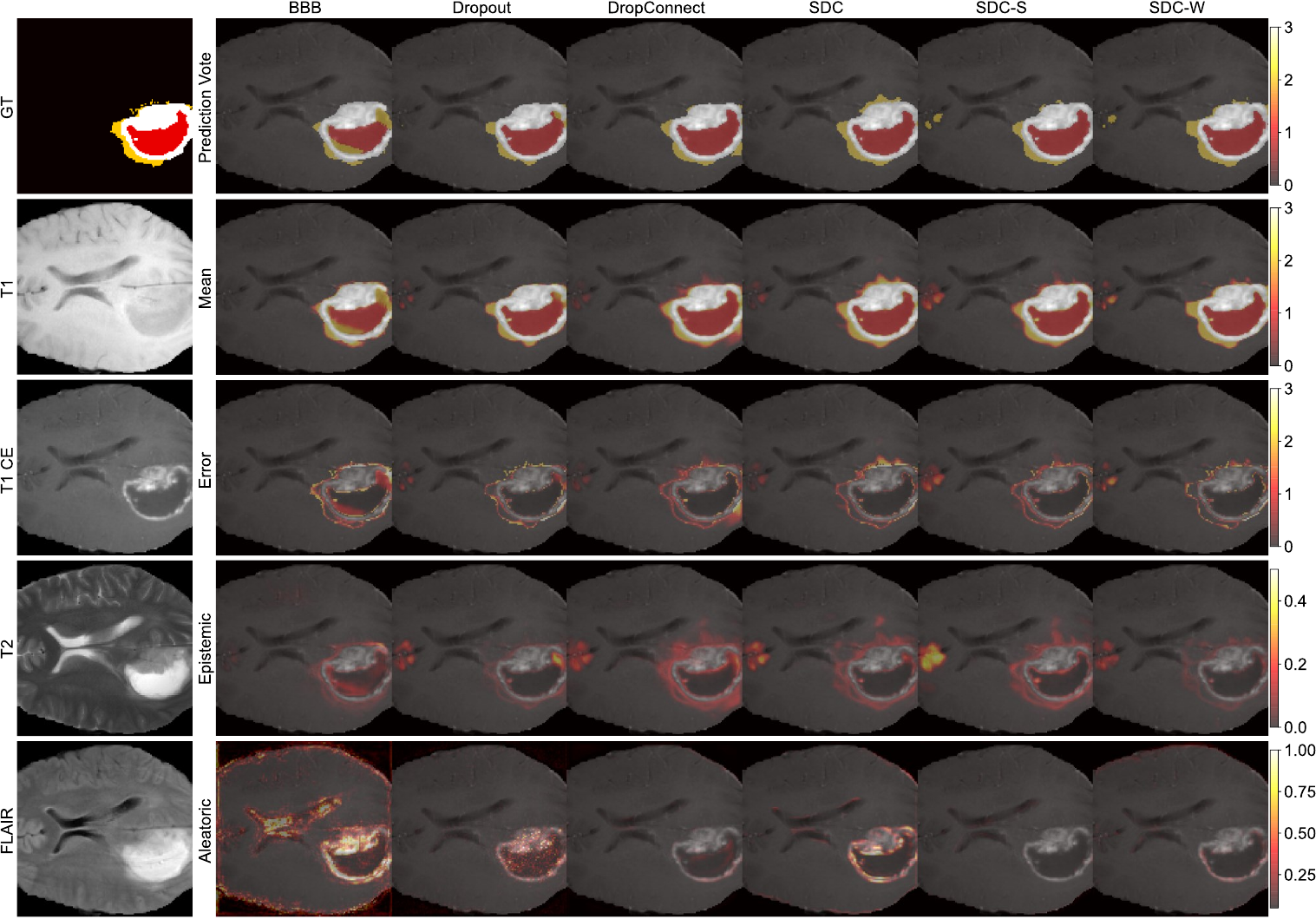}
\caption{An testing example of segmentation uncertainty with the leave out rate $p=0.05$. The left column shows ground truth and four MRI modalities. In the right part, each column shows prediction results using one method. The first row shows pixel-wise most popular votes. The second to fifth rows demonstrate pixel-wise average prediction, absolution error between ground truth and average predictions, pixel-wise epistemic uncertainty, and pixel-wise aleatoric uncertainty respectively.}
\label{fig4}
\end{figure}

Fig. \ref{fig4} represents an example of segmentation uncertain estimation. In terms of mean prediction results in the second row, consensus predictions lead to clearer edges and shapes, resulting in less uncertainties. In the last two rows, SDC-W's low pixel values reflect low uncertainty. Among all the six methods, SDC-W gives the best voting results. For example, the white spot pointed by the blue arrow can be clearly segmented in SDC-W's most popular voting results in the first row. For more examples, please see Supplementary Fig. 2 and 3.

\subsection*{Uncertain estimation on the brain metastases dataset for tumor origin site classification}
Finally, we used our own dataset to perform brain metastases origin site classification with network uncertainty estimation. Followed our previous work \cite{lyu2021transformer}, we upgraded the previously proposed classification framework with a capability of estimating the network uncertainty. Similar to what we did in the MNIST digit recognition task, here we also adopted the prediction class count, prediction popular voting class softmax possibility count to evaluate the network uncertainty.

\begin{figure}[t]
\centering
\includegraphics[width=\linewidth]{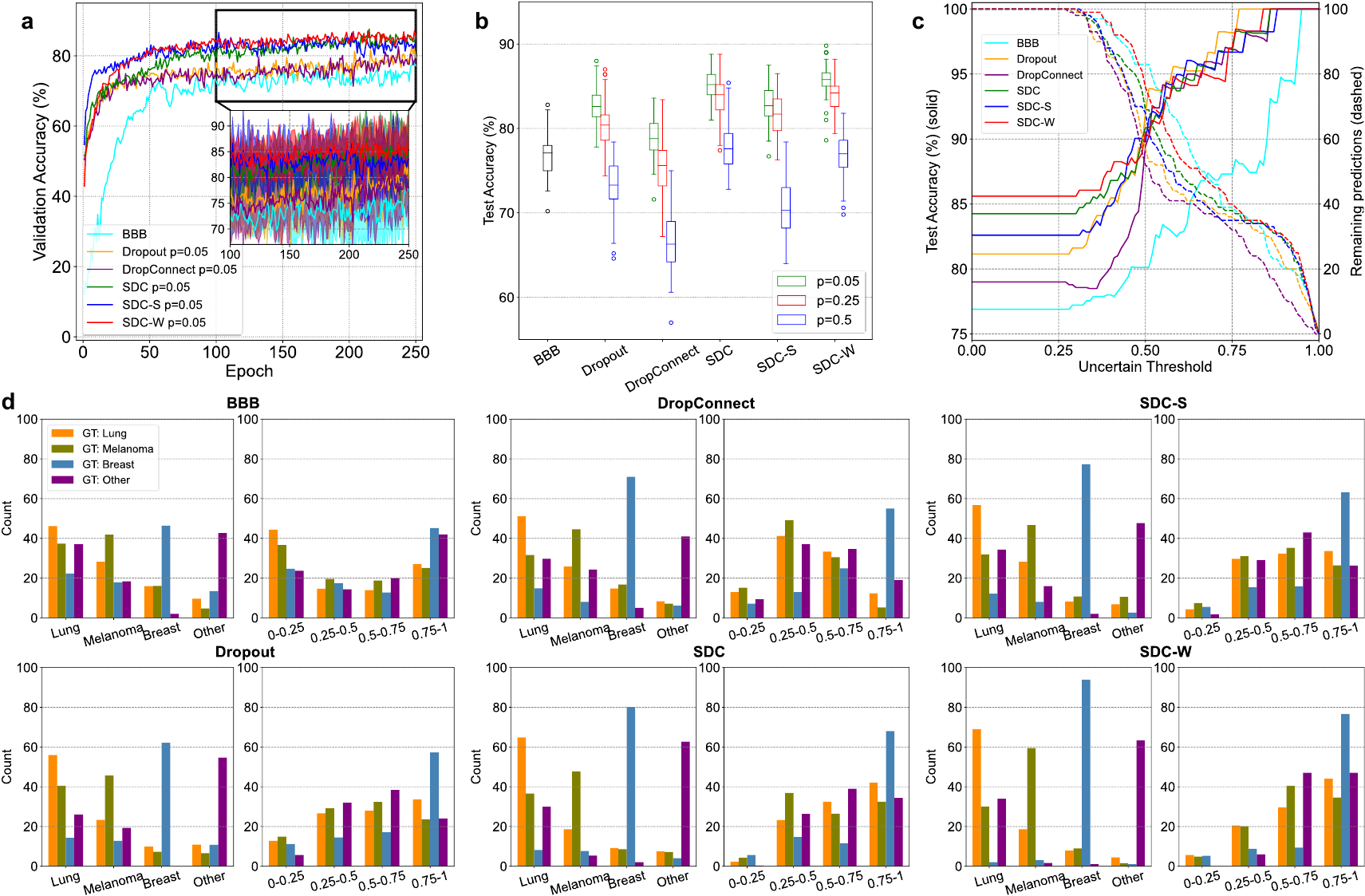}
\caption{Uncertainty estimation results on the brain metastases dataset for tumor origin site classification. a) The validation accuracy in the training process with the solid lines for average accuracy and bands for accuracy min-max range; b) Boxplot of whole tumor dice score on test data obtained using different methods; c) Double y-axis plot showing curves of prediction accuracy and curves of remaining predictions with respect to the uncertain threshold with $p=0.05$; d) Tumor origin site classification uncertainty testing results with $p=0.05$. In each case, the left subplot shows average each class prediction counts, and the right subplot shows average softmax possibility counts of prediction popular voting class.}
\label{fig5}
\end{figure}

In Fig. \ref{fig5}a-b, SDC-W gave the best classification results with the highest prediction accuracies in the validation and test phases. Compared with other methods with $p=0.05$, SDC-W substantially improved prediction accuracy by 11.7\%, 3.9\%, 8.7\%, 0.6\%, and 3.4\%, respectively. Fig. \ref{fig5}c demonstrates that the use of the network uncertainty helped avoid false predictions. The average softmax possibility of prediction popular voting class was used as the uncertainty threshold. When we gradually increase the uncertainty threshold, a growing number of prediction results with high uncertainty estimation will be rejected, meaning that a less number of predictions could be made. On the other hand, for the predictions the network makes, the accuracy should be increasingly higher. 

Fig. \ref{fig5}d represents the average prediction class counts and the average softmax possibility counts of prediction popular voting class. A higher average prediction class count indicates that prediction results tends to be more consistent in the test phase. When comparing all the six methods, it is clearly that SDC-W gives the highest average prediction class counts. In addition to the average prediction class count, we also considered the average softmax possibility counts of prediction class to evaluate the network uncertainty. The network becomes more confident about its predictions when more counts are in the right-most bin with the highest softmax possibility. Compared with BBB and Dropout, DropConnect, SDC, SDC-S and SDC-W generated more counts in those bins with a higher softmax possibility. Among all the methods, SDC-W seems the best, delivering the highest average prediction class counts and the most counts in relatively certain bins. For more examples, please see Supplementary Fig. 4 and 5.

\section*{Discussions}
The network uncertainty can be estimated by randomly weakening node connections in a neural network. Randomly weakening these links will generate stochastic fluctuations and information loss when information passes through the information processing workflow passage defined by the neural network. The stochastic information loss can be used to measure the network uncertainty. Suppose that there is a trained network which is very confident about its predictions, there must be sufficient information for the network to make predictions. Then, randomly leaving out a part of links would not change the network predictions dramatically because the network can utilize information redundancy to achieve performance robustness. On the other hand, for an unreliable network to make its predictions, the workflow would be much more fragile with respect to random connection dropouts, which should more likely change the predictions.

In this study, we have compared the six variational Bayesian inference methods, including our proposed SDC method and its two variants, on uncertainty estimation. Our experiments have been conducted for three tasks: one computer vision task for digit recognition, and two medical tasks for gliomas segmentation and metastases origin site classification respectively. Compared with the BBB, Dropout, and DropConnect methods, the proposed SDC method outperforms the others on all the three tasks, with higher accuracy and dice score as well as lower network uncertainty. Instead of directly discarding some connections of a node (DropConnect) or all connections of a node (Dropout), the proposed SDC method keeps node connects in principle while weakening them to different degrees by multiplying a random continuous variable such as $z^{ij}\sim U(0,1)$. Such a design keeps an integrity of the information flow within the network so that there is less information loss than Dropout and DropConnect. Consequently, at the Monte Carlo inference stage SDC would lead to less information loss, resulting in less network uncertainty.

Based on our generic SDC design, we have further modified it into two variants by limiting the range of multiplicative weight modulation factors to form SDC-S and SDC-W schemes. Compared to the generic SDC, weight modulation factors in SDC-S are smaller, indicating node connections are weakened more severely. On the other hand, weight modulation factors in SDC-W are closer to $1$ so that more information are kept. As shown in Fig. \ref{fig2}-\ref{fig4}, SDC-W shows less uncertainty than SDC and SDC-S, which is an interesting phenomenon and consistent to our insight behind the SDC approach (an analogy is that soft thresholding is better than hard thresholding).

We have also investigated the effect of the leave out rate $p$ on the network uncertainty. With the increment of $p$, more node connections are compromised, and there is more information loss in the information processing workflow, resulting in higher network uncertainty. Among the six methods compared in this study, SDC-W is the least affected one by the change of $p$, since there is generally the least information loss with SDC-W.

Network uncertainty estimation is not only useful but also necessary in medical applications. For segmentation tasks, as shown in Fig. \ref{fig4}, areas with high uncertainty measured by epistemic or aleatoric uncertainty are highly correlated with areas of false predictions. As a result, in real-world applications, radiologists should pay more attention to regions of high uncertainty, which helps mitigate false predictions. For classification tasks, Fig. \ref{fig5}c shows that higher prediction accuracy can be achieved by setting an appropriate uncertainty threshold to leave out some unreliable predictions. For those leave-out predictions, clinicians can perform classification aided by other methods such as biopsy \cite{vaidyanathan2019cancer} and bio-markers \cite{wu2015cancer}. Consequently, in reference to the network uncertainty, false predictions can be greatly alleviated. Clearly, this study shows a great potential of our SDC approach for network uncertainty estimation in future diagnosis and treatment planning, and improvement of healthcare outcomes and quality of life.


\bibliography{citation}

\section*{Acknowledgements}
This work was partly supported by National Institute of Biomedical Imaging and Bioengineering grants R01EB026646, and R01EB031102, National Cancer Institute grants R01CA233888, R21CA264772, P01CA207206 and P30CA012197, and National Heart, Lung, and Blood Institute grant R01HL151561.

\section*{Author Contributions}
CW defined the clinical problem.
QL and GW designed the network architecture.  
All authors designed the experiments.
QL conducted the experiments. 
All authors analyzed the results and revised the manuscript. 

\section*{Lead Contact}
Ge Wang, PhD (email:wangg6@rpi.edu).

\section*{Data and Code Availability}
MNIST and BraTS 2021 datasets are open access and can be downloaded online. To access whole-brain MRI data used in this paper, please contact the corresponding author Christopher T. Whitlow. All data will be available after the approval of materials transfer agreements (MTAs). Codes can be obtained at GitHub (\url{https://github.com/QingLyu0828/SoftDropConnect}).

\end{document}